\documentclass[12pt,reqno]{amsart}
\pdfoutput=1 
\usepackage{amsmath,bbm}
\usepackage{latexsym}
\usepackage{amsfonts}
\usepackage{amssymb}
\usepackage{color}
\usepackage{graphicx}
\usepackage{url}
\usepackage{enumerate}
\usepackage{tikz}
\usetikzlibrary{shadings, intersections, calc, plotmarks}
\usepackage{marginnote}
\usepackage{stackrel}

\usepackage{geometry}
\xdefinecolor{tumblue}     {RGB}{0,101,189}
\xdefinecolor{tumgreen}    {RGB}{162,173,  0}
\xdefinecolor{tumred}      {RGB}{229, 52, 24}
\xdefinecolor{tumivory}    {RGB}{218,215,203}
\xdefinecolor{tumorange}   {RGB}{227,114, 34}
\xdefinecolor{tumlightblue}{RGB}{152,198,234}

\newtheorem{proposition}{Proposition}
\newtheorem{theorem}{Theorem}
\newtheorem*{theorem*}{Theorem}

\newtheorem*{corollary*}{Corollary}

\newtheorem{definition}{Definition}
\newtheorem{conjecture}{Conjecture}

\newcommand{\R}{\mathbbm{R}}
\newcommand{\C}{\mathbbm{C}}
\newcommand{\N}{\mathbbm{N}}

\newcommand{\F}{\mathbbm{F}}
\newcommand{\Q}{\mathbbm{Q}}

\newcommand{\1}{\mathbbm{1}}
\newcommand{\Red}{\mathrm{Red}}

\def\>{{\rangle}}
\def\<{{\langle}}

\newcommand{\be}{\begin{equation}}
\newcommand{\ee}{\end{equation}}
\newcommand{\bea}{\begin{eqnarray}}
\newcommand{\eea}{\end{eqnarray}}

\newcommand{\tr}[1]{\mathrm{tr}\left[#1\right]} %trace

\begin{document}

\title[On the set of reduced states]{On the set of reduced states of translation invariant, infinite quantum systems }

\author[Blakaj]{Vjosa Blakaj$^{1,2}$}
\email{vjosa.blakaj@tum.de}
\author[Wolf]{Michael M. Wolf$^{1,2}$}
\email{m.wolf@tum.de}
\address{$^1$ Department of Mathematics, Technical University of Munich}
\address{$^2$ Munich Center for Quantum
Science and Technology (MCQST),  M\"unchen, Germany}

\begin{abstract} 
The set of two-body reduced states of translation invariant, infinite quantum spin chains can be approximated from inside and outside using matrix product states and marginals of finite systems, respectively. These lead to hierarchies of algebraic approximations that become tight only in the limit of infinitely many auxiliary variables. We show that this is necessarily so for any algebraic ansatz by proving that the set of reduced states is not semialgebraic. We also provide evidence that additional elementary transcendental functions cannot lead to a finitary description. 
\end{abstract}

\maketitle
\tableofcontents

%%%%%%%%%%%%%%%%%%%%%%%%%%%%%%%%%%%%%%%%
\section{Introduction}\label{sec:intro}
%%%%%%%%%%%%%%%%%%%%%%%%%%%%%%%%%%%%%%%%

Quantum correlations become non-trivially constrained when considering parts of a larger system that obeys symmetry rules. This fact is exhibited most prominently by the \emph{monogamy of entanglement}. More generally, it is manifestly inherent in all \emph{quantum marginal problems} that concern the consistency of reduced density matrices with global, often symmetry-based, constraints. A central motivation for studying these problems is the fact that ground-state energies in quantum chemistry \cite{coleman2000reduced} and condensed matter physics \cite{VerstraeteCirac06} depend solely on the reduced density matrices under global symmetry constraints. Also, phase transitions and symmetry breaking are reflected in the geometry of the set of reduced density matrices \cite{Zauner_2016}.

\vspace{3pt}

In this paper, we consider the set of reduced two-body density matrices that arise in translationally invariant, infinite quantum spin chains. In these systems, the ground state energy density of a nearest-neighbor interaction Hamiltonian could in principle be obtained by optimizing over all admissible reduced states. However, even in the case of two-dimensional local Hilbert spaces, this set is not known exactly. There are inner approximations based on mean-field and matrix-product techniques \cite{VerstraeteCirac06} and outer approximations based on relaxations of constraints \cite{kull2022lower,haim2020variationalcorrelations,Baumgratz_2012}. When unraveling these approximations, which are usually expressed directly as algorithms for bounding ground state energy densities, they all correspond to \emph{semialgebraic} sets of reduced density matrices. That is, sets that can be described by a finite number of algebraic (in-)equalities. Our aim is to prove that the set of interest, i.e., the set of admissible two-body reduced density matrices, is itself not semialgebraic. In the last section, we show that the situation is most likely even worse (or more interesting) by providing evidence that additional elementary transcendental functions are still not sufficient to allow an explicit description by finite means. 

\vspace{3pt}

Our work is inspired by an argument of Fannes, Nachtergaele, and Werner. In \cite{FCS} they show that the ground state of the antiferromagnetic spin 1/2 Heisenberg chain cannot be a $C^*$-finitely correlated state: its energy is transcendental while the algebraic ansatz can only lead to algebraic energies. This argument holds for any parameter-free algebraic ansatz but fails as soon as transcendental parameters are allowed. For this reason, we invoke transcendental sets and functions rather than numbers.

%%%%%%%%%%%%%%%%%%%%%%%%%%%%%%%%%%%%%%%%
\section{(Semi-)algebraic preliminaries}\label{sec:prelim}
%%%%%%%%%%%%%%%%%%%%%%%%%%%%%%%%%%%%%%%%

We begin with reviewing some mathematical concepts and introducing terms and notations.

A function $f:I\rightarrow\R$ on an interval $I\subseteq\R$ is called \emph{algebraic} over a subfield $\F\subseteq\R$ if there is a polynomial $p\in\F[y,x]$ and an interval $J\subseteq\R$ s.t. $y=f(x) \Leftrightarrow p(y,x)=0$ holds for all $(x,y)\in I\times J$. A function that is not piecewise algebraic over $\F$ is called \emph{transcendental} over $\F$. We emphasize the field $\F$ since these notions are often understood with $\F=\Q$ in mind, but we will need $\F=\R$. 

A set $A \subseteq \mathbb{R}^n$ is called \textit{semialgebraic} over a subring $R\subseteq\R$ if it can be defined by a finite number of polynomial equalities and inequalities with coefficients in $R$. A powerful tool for showing that a set is semialgebraic is the Tarski-Seidenberg theorem, which enables the use of quantifiers without leaving the semialgebraic world:

\begin{theorem}[Tarski-Seidenberg quantifier elimination (cf.\cite{Basu})]\label{thm:TarskiSeidenberg}
Let $R\subseteq\R$ be a subring and $\{p_j(v,x)\}_{j=1}^l$  a finite set of polynomial equalities and inequalities with coefficients in $R$ and variables $(v,x) \in \R^k \times \R^n$. If $\phi(v,x)$ is a Boolean combination (using $\wedge,\vee,\neg$) of the $p_j$'s and 
\begin{equation}
    \Psi(x) := \big(Q_1 v_1 \ldots Q_k v_k : \phi(v,x) \big), \,\,\, Q_i \in \{ \exists, \forall \},
\end{equation}
then there exists a quantifier-free formula $\psi(x)$ consisting of a Boolean combination of finitely many polynomial (in-)equalities with coefficients in $R$, s.t.
\begin{equation}
    \forall x: \,\, \big( \psi(x) \Leftrightarrow \Psi(x) \big).
\end{equation}
Moreover, there exists an effective algorithm that constructs  $\psi$ from $\Psi$.
\end{theorem}
In our case, $\R^n$ will be a real vector space of complex Hermitian matrices, represented in some Hermitian operator basis. The choice of the basis does not matter as long as $R=\R$ is considered.

\vspace{3pt}

A function is called \emph{semialgebraic} over $R$ if its graph is a semialgebraic set. If $f:\R\supseteq I\rightarrow \R$ is semialgebraic over $\R$, then it is piecewise algebraic over $\R$ (cf. Sec.2.5.2 in \cite{Basu}).

%A number $\alpha \in \C$ is said to be \textit{algebraic} if it is the root of a non-zero polynomial $p \in \Q[x]$ of finite degree. We denote by $\overline{\Q}$ the field of algebraic numbers. A set of numbers $\{z_1,\ldots, z_k\}$ is called \textit{algebraically independent} if the zero polynomial is the only polynomial $p \in \overline{\Q}[x_1, \ldots, x_k]$ that satisfies $p(z_1,\ldots,z_k)=0$.

%%%%%%%%%%%%%%%%%%%%%%%%%%%%%%%%%%%%%%%%
\section{Reduced density matrices}\label{sec:result}
%%%%%%%%%%%%%%%%%%%%%%%%%%%%%%%%%%%%%%%%

Consider a multipartite quantum system of spins aligned on a one-dimensional chain such that each spin is assigned a Hilbert space $\C^d$. Our interest lies in the reduced density matrices of neighboring spins that are admissible under the constraint of global translation symmetry. 
\begin{definition} Let $\Red_d$ denote the subset of density operators on $\C^d\otimes\C^d$ s.t. for every $N\in\N$ there is a density operator $\rho^{(N+1)}$ on $(\C^d)^{\otimes(N+1)}$ with the property that all its successive bipartite reduced states are equal to $\rho$, i.e.,
\begin{equation}
    {\rm tr}_{\neg (i,i+1)}\big[\rho^{(N+1)}\big]=\rho\qquad \forall i\in\{1,\ldots,N\}.\label{eq:Redconstraints}
\end{equation}
\end{definition}
More informally, $\rho\in\Red_d$ if and only if $\rho$ can be extended in a translational invariant way to an infinite quantum spin chain.
\begin{proposition} $\Red_d$ is convex and compact.\label{prop:compact}
\end{proposition}
\begin{proof}
    Convexity follows readily from the convexity of the set of density matrices and the linearity of the partial trace. Moreover, boundedness is inherited from the set of all density operators. In order to see closedness, we write $\rho_{i,i+1}^{(N+1)}:={\rm tr}_{\neg (i,i+1)}\big[\rho^{(N+1)}\big]$ and define
    \begin{equation}
        f\left(\rho^{(N+1)}\right):=\sum_{i,j=1}^N\big\|\rho_{i,i+1}^{(N+1)}-\rho_{j,j+1}^{(N+1)}\big\|
    \end{equation} on the set of $(N+1)$-partite density operators. Since $f$ is continuous, the set $S_{N+1}:=f^{-1}\big(\{0\}\big)$ is closed and thus compact. Hence, $R_{N+1}:={\rm tr}_{\neg (1,2)}[S_{N+1}]$ is compact as the image of a compact set under a continuous map. Finally, since $R_{N+1}$ is the set of bipartite density operators that can be extended to an $(N+1)$-partite state with equal successive bipartite marginals, we see that $\Red_d=\bigcap_{N\in\N} R_{N+1}$ is compact as it is an intersection of compact sets.
\end{proof}

\vspace{2pt}

Given a two-body Hamiltonian described by a Hermitian operator $h$ acting on $\C^d\otimes\C^d$, we define $h_{i,i+1}:=\1^{\otimes (i-1)}\otimes h\otimes\1^{\otimes (N-i-1)}$ acting on $(\C^d)^{\otimes N}$. The corresponding  \emph{ground state energy density} in the \emph{thermodynamic limit} ($N\rightarrow\infty$) is then 
\begin{equation}\label{eq:epsdef}
    \epsilon:=\lim_{N\rightarrow\infty} \frac{1}{N} \;\inf_{\rho^{(N)}}\;\sum_{i=1}^{N-1} \tr{\rho^{(N)}h_{i,i+1}},
\end{equation}
where the infimum is taken over all $N$-partite density operators $\rho^{(N)}$.

\vspace{2pt}

The following is well known \cite{VerstraeteCirac06}, but we provide a proof for a more self-consistent and coherent presentation. 
\begin{proposition}
    For every two-body Hamiltonian $h$ the corresponding ground state energy density in the thermodynamic limit is given by 
    \begin{equation}
        \epsilon=\min_{\rho\in\Red_d} \tr{\rho h}.\label{eq:epsRedd}
    \end{equation}
\end{proposition}
\begin{proof} First note that compactness of $\Red_d$ (Prop.\ref{prop:compact}) guarantees the existence of a minimizer.
    The r.h.s. of Eq.(\ref{eq:epsRedd}) can be seen to be an upper bound on $\epsilon$ by restricting the infimum in Eq.(\ref{eq:epsdef}) to density operators with equal bipartite marginals. This leads to
    \begin{equation*}
        \epsilon\leq \lim_{N\rightarrow \infty} \frac{N-1}{N} \min_{\rho\in\Red_d} \tr{\rho h}=\min_{\rho\in\Red_d} \tr{\rho h}.
    \end{equation*}
    In order to obtain an inequality in the other direction, we add an additional interaction between the $N$'th and the first particle so that the resulting overall Hamiltonian becomes cyclic on a ring of size $N$. Since the change in energy  is at most $\|h\|$, we can bound
    \begin{eqnarray}\label{eq:iuhgeff}
        \epsilon&\geq& \lim_{N\rightarrow\infty} \frac{1}{N}\left[-\|h\|+\min_{\rho^{(N)}}\sum_{i=1}^{N}\tr{\rho^{(N)}h_{i,i+1}}\right]\\
        &=& \lim_{N\rightarrow\infty} \frac{1}{N}\left[-\|h\|+N \min_{\rho\in\Red_d}\tr{\rho h}\right] = \min_{\rho\in\Red_d}\tr{\rho h}.\label{eq:kwefw}
    \end{eqnarray}
    Here, we have identified the $N+1$'st with the first site in Eq.(\ref{eq:iuhgeff}) and used for Eq.(\ref{eq:kwefw}) that the energy of a translational invariant Hamiltonian with periodic boundary condition is minimized by a translational invariant state. That is, in particular, by a state with equal bipartite marginals.
\end{proof}

\begin{theorem}\label{thm:main} For $d\geq 2$, the set $\Red_d$ is not semialgebraic over $\R$.
\end{theorem}
\begin{proof}
    We will show this by proof of contradiction, starting from the assumption that $\Red_d$ is semialgebraic. On an interval $I\subseteq\R$ consider a mapping $I\ni\gamma\mapsto h(\gamma)$   into the set of Hermitian matrices on $\C^d\otimes \C^d$ with the property that each matrix entry of $h(\gamma)$ is a polynomial function of $\gamma$. We may, for instance, consider a simple affine relation of the form $h(\gamma)=h_0+\gamma h_1$. 
    
    Regarding $h(\gamma)$ as a parameter-dependent interaction Hamiltonian, the graph of the corresponding ground state energy density $\epsilon(\gamma)$ on the interval $I$ (a semialgebraic set) can be expressed as 
    \begin{subequations}\label{eq:semiepsilon}
        \begin{align}
        \big\{\big(\epsilon(\gamma),\gamma\big)\;\big|\;\gamma\in I\big\}\ = \ \big\{(E,\gamma)\;|\;& E\in\R,\;\gamma\in I,\\
        &\exists \rho\in\Red_d:\ \tr{\rho h(\gamma)}=E,\\
        &\forall \sigma\in\Red_d:\ \tr{\sigma h(\gamma)}\geq E\ \big\}.\qquad
    \end{align}
    \end{subequations}
    Under the assumptions that $h$ is polynomial and $\Red_d$ semialgebraic over $\R$, the Tarski-Seidenberg theorem guarantees that the set defined in Eq.(\ref{eq:semiepsilon}) is semialgebraic since all equations are polynomial and all quantifiers run over semialgebraic sets. The fact that the quantifiers in Eq.(\ref{eq:semiepsilon}) are not at the beginning is not an obstacle since every first-order formula can be brought to \emph{prenex normal form}, which is used in Thm.\ref{thm:TarskiSeidenberg}.
    
    Consequently, $I\ni\gamma\mapsto\epsilon(\gamma)$ is a semialgebraic function and therefore piecewise algebraic over $\R$.  Hence, the proof is completed by any example whose ground state energy density is transcendental over $\R$ (despite the fact that $\gamma\mapsto h(\gamma)$ is polynomial). The next section will provide such an example for the case $d=2$. This also covers the case of higher dimensions by simply embedding smaller Hilbert spaces into larger ones.
\end{proof}
% \textit{Comparison to the classical case}

%%%%%%%%%%%%%%%%%%%%%%%%%%%%%%%%%%%%%%%%
\section{Transcendental ground state energy density}\label{sec:XY}
%%%%%%%%%%%%%%%%%%%%%%%%%%%%%%%%%%%%%%%%
As a special example for the completion of the proof of Thm.\ref{thm:main}, we consider the anisotropic XY-model, which is specified by the two-body Hamiltonian 
\begin{equation}
    h(\gamma):=(1-\gamma) \sigma_x\otimes\sigma_x +(1+\gamma)\sigma_y\otimes\sigma_y,\quad \gamma\in(-1,1).
\end{equation}

\begin{theorem} In the thermodynamic limit, the ground state energy density  $\epsilon(\gamma)$ of the anisotropic XY-model is an analytic function that is transcendental over $\R$. 
\end{theorem}
\begin{proof}
The ground state energy density in the thermodynamic limit was derived in \cite{Lieb1961} and shown to be
\begin{equation}\label{eq:egamma}
    \epsilon(\gamma)=-\frac{1}{4\pi}\int_0^{\pi/2}\Big[1-\underbrace{(1-\gamma^2)}_{=:z^2} \sin^2(k)\Big]^{1/2} dk =:-\frac{1}{4\pi}E(z),
\end{equation}
where $E(z)$ is the \emph{complete elliptic integral of the second kind}. Suppose that $\epsilon$ is algebraic over $\R$ in some neighborhood. Since the set of algebraic functions is closed under composition, the same would be true for $E$.   
The function $E$ is known to be an analytic solution to the \emph{hypergeometric differential equation}
\begin{equation}\label{eq:hgeq}
    z(1-z)E''+\big(c-(a+b+1)z\big)E'=ab E,
\end{equation} for $(a,b,c)=(-1/2,1/2,1)$ (see 17.3.10 and 15.5.1 in \cite{AbraSteg72}). Due to the analyticity of $E$, we can extend the view to the complex plane such that a supposed polynomial relation over $\R$ would imply one over $\C$.

\vspace{2pt}

Differential equations of the kind of Eq.(\ref{eq:hgeq}) have an algebraic solution if and only if they have a finite monodromy group (cf. Thm. 5.11 in \cite{haraoka2020linear}). For  Eq.(\ref{eq:hgeq}), these cases were completely characterized by Schwarz in \cite{Schwarz1873}, who determined a complete list of corresponding triples $(a,b,c)$ (summarized e.g. in Tab.2 of \cite{BOD2012541}), which does not contain  $(a,b,c)=(-1/2,1/2,1)$. Consequently, $E$ and therefore also $\epsilon$ are transcendental functions over $\R$. 
\end{proof}

%Due to analyticity of $E$, we can extend the viewpoint to the complex plane

%%%%%%%%%%%%%%%%%%%%%%%%%%%%%%%%%%%%%%%%
\section{Extensions and implications}\label{sec:impl}
%%%%%%%%%%%%%%%%%%%%%%%%%%%%%%%%%%%%%%%%

The fact that $\Red_d$ is not semialgebraic over $\R$ implies that no algebraic ansatz  (with or without transcendental parameters) can describe the set exactly without invoking a limit of infinitely many variables. The best-known example is the set of matrix product states with increasing bond dimension. This raises the question of whether supplementing algebraic methods with transcendental functions could enable a finitary, explicit description of $\Red_d$. In \cite{fawzi2023entropy}, for instance, entropy constraints, which are known to be transcendental \cite{Blakaj2023, blakaj2023transcendental}, are used in addition.

\vspace{2pt}

Evidence against such finitary descriptions we can see from two sides: first, functions related to elliptic integrals (as in Eq.(\ref{eq:egamma})) can often be shown to be non-elementary \cite{TopoGalois,Kasper} in the sense that they can not be expressed in terms of algebraic and elementary transcendental functions---neither explicitly nor implicitly.
At the moment, however, we do not know exactly what the implications are for $\Red_d$. We will therefore pursue a second line of thought more closely.

\vspace{2pt}

We want to allow the additional use of the exponential function $\exp$ (and thereby implicitly also of $\ln$). To this end, denote by $\R_{\exp{}}:=(\R,+,\cdot,\exp{},<,0,1)$ the ordered field of real numbers with exponentiation, and call a subset $A\subseteq\R^n$ \emph{definable} in the first-order language of $\R_{\exp{}}$ if $a\in A \Leftrightarrow \Phi(a)$ for some first-order formula $\Phi$. That is, there is a $k\in\N$ and a Boolean combination $\phi$ of finitely many (in-)equalities of polynomial\footnote{with algebraic coefficients} and exponential functions in $n+k$ real variables, s.t.
    \begin{equation}
        \Phi(a)=Q_1 b_1 \cdots Q_k b_k\ \phi(a_1,\ldots,a_n,b_1,\ldots,b_k),
    \end{equation}
where each $Q_i\in\{\exists,\forall\}$ is a quantifier. Unlike for $\R$ there is no quantifier elimination for  $\R_{\exp{}}$ \cite{DENDRIES198497}. However, in \cite{Macintyre1996-MACOTD-3}  Macintyre and Wilkie proved that the first-order theory of $\R_{\exp{}}$ is decidable conditioned on the validity of the following:

\begin{conjecture}[Schanuel's conjecture (cf. Chap.21 in \cite{MurtyRath})] If $z_1,\ldots,z_n \in \C$ are linearly independent over $\Q$, then $\{z_1,\ldots,z_n,\exp{z_1},\ldots,\exp{z_n}\}$ contains at least $n$ algebraically independent numbers.\footnote{A set of numbers is called \emph{algebraically independent} if there is no non-zero multivariate polynomial with coefficients in $\Q$ that has these numbers as roots. }
\end{conjecture}
Schanuel's conjecture can be considered \emph{the} central conjecture of transcendental number theory. The best-known proven special case is the one where all $z_i$'s are algebraic numbers. Then this becomes the content of the Lindemann-Weierstrass theorem. 

\vspace{2pt}

Returning to the set of reduced density matrices, we obtain the following: 

\begin{theorem}\label{thm:Schanuel}
    Assuming Schanuel's conjecture, there exists a $d\in\N$ such that the set $\Red_d$ is not definable in the first-order language of $\R_{\exp{}}$. 
\end{theorem}

\begin{proof} By the results of \cite{1Dundecidablegap,undecidablegap} there is a $d\in\N$ and a family of interaction Hamiltonians $h$ on $\C^d\otimes\C^d$ with matrix entries in $\Q[\sqrt{2}]$ (and thus expressible in $\R_{\exp{}}$) such that the ground state energy density problem
\begin{equation}\label{eq:undecgsed}
    \forall \rho\in\Red_d:\;\tr{h\rho}>0
\end{equation}
is undecidable. Assuming that $\Red_d$ was definable in the first-order language of $\R_{\exp{}}$, then Eq.(\ref{eq:undecgsed}) would be expressible as a sentence in the first-order language of $\R_{\exp{}}$. So according to the result of  Macintyre and Wilkie \cite{Macintyre1996-MACOTD-3}, Eq.(\ref{eq:undecgsed}) would be decidable if we assume Schanuel's conjecture.
\end{proof}

As shown in \cite{BERARDUCCI200443},  Schanuel's conjecture in Thm.\ref{thm:Schanuel} can be replaced by the so-called \emph{transfer conjecture}, which also implies decidability of the first order theory of $\R_{\exp{}}$. 
Moreover, there are closely related results that do not rely on any unproven conjecture. For instance, it was proven in \cite{MCCALLUM201216}  that sentences in the first-order theory of $\R_{\rm trans}$, for a transcendental function ${\rm trans}\in\{\exp,\ln,\arctan,\ldots\}$, are decidable if only the outermost quantified
variable occurs in the transcendental function.

\vspace{0.3cm}

\noindent {\it Acknowledgments.} VB thanks Bruno Nachtergaele, J. Ignacio Cirac, Georgios Styliaris, Chokri Manai, Amando Young, and Esther Cruz Rico for insightful discussions. 
This work has been supported by the Deutsche Forschungsgemeinschaft (DFG, German Research Foundation) under Germany's Excellence Strategy EXC-2111 390814868, the SFB/Transregio TRR 352 – Project-ID 470903074, and the International Max Planck Research School for Quantum Science and Technology at the Max-Planck-Institute of Quantum Optics.

\bibliographystyle{halpha}
\bibliography{Bakery}{}

\end{document}